\def\rn{\noindent\parshape 2 0truecm 8.5truecm 0.3truecm 8.2truecm}
\def\rn{}
\def\nn#1 #2{#2. #1}				
\def\nnn#1 #2 #3{#2. #3. #1}			
\def\nnnn#1 #2 #3 #4{#2. #3. #4 #1}		
\def\nnnnn#1 #2 #3 #4 #5{#2. #3. #4 #5. #1}	

\def\rf#1;#2;#3;#4;#5 {{\frenchspacing\par\rn#1, #3 {\bf #4}, #5 (#2). \par}}
\def\rg#1;#2;#3;#4;#5;#6 {{\frenchspacing\par\rn#1, #3 {\bf #4}, #5 (#2). \par}}
\def\rfbook#1;#2;#3;#4;#5 {{\frenchspacing\par\rn#1, {\it #3} (#5, #4, #2).\par}}
\def\rfprep#1;#2;#3 {{\par\frenchspacing\rn#1, #3 (#2).\par}}

\def\beq#1{\begin{equation}\label{#1}}
\def\eeq{\end{equation}}
\def\beqa#1{\begin{eqnarray}\label{#1}}
\def\eeqa{\end{eqnarray}}

\def\spose#1{\hbox to 0pt{#1\hss}}
\def\simlt{\mathrel{\spose{\lower 3pt\hbox{$\mathchar"218$}}
     \raise 2.0pt\hbox{$\mathchar"13C$}}}
\def\simgt{\mathrel{\spose{\lower 3pt\hbox{$\mathchar"218$}}
     \raise 2.0pt\hbox{$\mathchar"13E$}}}
\def\simpropto{\mathrel{\spose{\lower 3pt\hbox{$\mathchar"218$}}
     \raise 2.0pt\hbox{$\propto$}}}

\def\ed{\end{document}}

\def\draft{
}

\def\k{{\bf k}}
\def\x{{\bf x}}

\def\x{{\bf x}}

\documentstyle[prd,aps,epsf]{revtex}
\draft
\begin{document}
\twocolumn[\hsize\textwidth\columnwidth\hsize\csname@twocolumnfalse\endcsname



\title{Non-Gaussianities in models with a varying inflaton decay rate}

\author{Matias Zaldarriaga}

\address{Department of Astronomy and Department of Physics, Harvard University, MA 02138}

\date{\today. To be submitted to Phys. Rev. D.}

\maketitle 

\begin{abstract}
We consider the expected level of primordial  non-Gaussianities in models in which density perturbations are produced by spatial fluctuations in the decay rate of the inflaton. We consider both the  non-Gaussianities resulting from the self-couplings of the field that controls the decay rate as well as from the non-linear relation between field and curvature perturbations. We show that  in these scenario non-Gaussianities are of the ``local" form, {\it ie} well described be the ansatz 
${\cal R} ={\cal R}_{g} + f_{NL}^{{\cal R}} ({\cal R}_{g}^2 - \langle {\cal R}_{g}^2 \rangle )$. This is a consequence of the fact that they were created when modes were already outside the horizon. We show that $ f_{NL}^{{\cal R}} $ is naturally of order a few in these models, much larger than what is expected in the standard one field models of inflation ($ f_{NL}^{{\cal R}} \sim 10^{-2}$) and possibly accessible to observations.  

\end{abstract}
\bigskip
]


\section{Introduction} 

It is safe to say that the era of  ``precision cosmology" is upon us. The latest CMB anisotropy measurements from WMAP \cite{bennet} have underscored how successful our standard cosmological model is. Many of the parameters describing the matter content of the universe are now firmly established \cite{spergel}. Moreover the conclusion that the structure we observe in the universe grew from seeds that were in place much before recombination seems almost  unavoidable \cite{pieris}.


The leading scenario for explaining how these initial seeds were created is inflation. In this picture 
the universe went through a period of accelerated expansion very early on. During inflation the energy density is dominated by the potential energy of a slowly rolling scalar field, the inflaton. During this period quantum fluctuations in the inflaton field were stretched outside the horizon and became the seeds for structure. The predictions from this scenario have been worked out to great detail in the last few decades (see for example \cite{linde,liddlelyth} and references therein). Basically most models of inflation predict a scale invariant spectrum of Gaussian fluctuations of the curvature. A stochastic background of gravitational waves is also predicted, although its level is quite uncertain. 

The Gaussianity of the perturbations in the standard picture are a consequence of the slow-roll conditions on the inflaton potential. These conditions are required for the field to be slowly rolling and the universe to be inflating. The required flatness of the potential implies that non-Gaussianities are very small, probably below what could be observable (see \cite{maldacena} and references therein). If the inflaton field has several components it is possible to construct models with larger non-Gaussianities even in the adiabatic component. Basically large couplings are allowed for the ``isocurvature" directions which if suitably coupled to the adiabatic direction could lead to non-Gaussianities in that direction as well (eg. \cite{bernardeau1,bernardeau2,bartolo2}).


There are basically two modifications of the standard inflationary scenario  in which the curvature perturbations  we observe today were not created during  inflaton  but later.  These are the ``curvaton" scenario \cite{mollerach,linde2,lyth,moroi,enqvist,bartolo} and the scenario proposed in \cite{dgz} (see also \cite{dgz2,kofman}). Both require the existence of other light fields during inflation. In the ``curvaton" model one of these fields eventually comes to dominate or at least be a significant contributor to the energy density and reheats the universe again. The curvature perturbations we observe  are a result of the fluctuations in the curvaton field as opposed to the inflaton. Fluctuations in the curvaton were still produced during inflation by the standard mechanism but they get converted into curvature perturbations when the curvaton becomes dominant.  

The scenario presented in \cite{dgz} is different. At the end of inflation the energy stored in the inflaton has to be converted into normal particles, reheating the universe and starting the standard phase of the hot big bang. In \cite{dgz} we suggested that if the efficiency with which the inflaton reheats the universe, its decay rate $\Gamma$ varied in space, density perturbations would be generated during reheating independently of those generated by the standard inflationary  mechanism. If two different regions of the universe had different $\Gamma$s then effectively the inflaton would decay into radiation first in one region and then in the other. During the time one region is filled with radiation while the other one is not, the universe expands at a different rate in each region, resulting in density perturbations when reheating is finished. In a similar spirit, the in-homogeneous freeze out of an unstable  particle whose mass and decay rate are determined by a light scalar field  and that briefly dominates the energy density would also lead to density perturbations \cite{dgz2}. 

The motivation for these new scenarios is that in general the decay rate of the inflaton as well as the masses and decay rates of particles are determined by the expectation values of scalar fields. If those scalar fields were light during inflation they fluctuated, which lead to  density perturbations. 


In this paper we will analyze under what conditions the fluctuations in the scenario of \cite{dgz} are Gaussian. For example, the potential for the fields that determine the inflaton decay rate do not have to satisfy any slow-roll conditions so one might expect their fluctuations to be significantly non-Gaussian. We will find that generically we predict non-Gaussianities that are larger than those predicted by the standard inflationary model although within the observational bounds. Moreover these additional non-Gaussianities could be accesible to future experiments. There are two avenues for searching for primordial non-Gaussianities observationally, studies of the CMB \cite{komsper} and studies of Large Scale structure as traced say by galaxies or weak lensing (see \cite{romanreview} for a comprehensive review of non-Gaussinanity  in the context of Large Scale Structure).  

The paper is organized as follows. In \S \ref{s1}  we summarize how the curvature perturbations are generated in this scenario. In \S \ref{s2} we discuss the different sources of non-Gaussianities and finally in \S \ref{s3} we make contact with observations.

\section{Density perturbations in the new scenario}\label{s1}

 The decay rate of the inflaton (which we call $S$) can be written schematically as,
\begin{equation}
\Gamma = m_S\ g^2 \ K ,
\end{equation}
where $m_S$ is the mass of the inflaton, $g$ its coupling constant and $K$ are kinematic suppression factors of the form $(1- m_d^2 / m_S^2)$ where $m_d$ stands for the masses of the decay products which are also functions of the scalar fields. All three quantities, $m_S$,  $g$ and $K$ are functions of the scalar fields in the theory so clearly there are a wide rage of possible specific models in which $\Gamma$ has a different dependance on the fields

During inflation the light fields $\phi_i$ have expectation values $\bar \phi_i$ around which they fluctuate.  To keep our discussion generic we will express the decay rate as an expansion in powers of those fluctuations, 
\begin{equation}
\Gamma = \Gamma_0 +{ \partial \Gamma \over \partial \phi_i} \delta \phi_i + {1\over 2} { \partial^2 \Gamma \over \partial \phi_i \partial \phi_j} \delta \phi_i \delta \phi_j + \cdots 
\label{gammaexp}
\end{equation}

In \cite{dgz} it was shown that to linear order in the decay rate fluctuations ${\delta\Gamma / \Gamma}$ curvature fluctuations $\cal R$ are generated.  We found, 
 \begin{equation}
{\cal R} = - \alpha {\delta\Gamma \over \bar  \Gamma}
\end{equation}
where the constant $\alpha$ depends on the ratio of $\bar \Gamma$ to the Hubble expansion at the end of inflation $(H)$. As $\bar \Gamma/H$ increases $\alpha$ decreases. It is clear for example  that in the limit $\Gamma / H >> 1$ reheating happens almost instantaneously in all regions regardless of the value of $\Gamma$ and thus the resulting density perturbations will be very small.  In \cite{dgz} it was proved that in the limit when $\bar \Gamma << H$ 
 \begin{equation}
{\cal R} = -{1 \over 6}  {\delta \Gamma \over \Gamma}.
\end{equation}
We will now present a different derivation of the same result that will allow us to extend our treatment to investigate non-Gaussianities in a later section. 

We first consider reheating in a homogeneous universe, ie with a constant $\Gamma$.   We will treat the inflation as non-relativistic matter when it oscillates around the  minimum of its potential after inflation ends.  We will call $\epsilon _m$ the energy density in the oscillating inflaton.  During  reheating, the energy density in the inflaton  is converted into radiation (with energy density $\epsilon _r$). The background metric is that of a flat FRW universe, $ds^2=-dt^2+a^2(t)dx^2$, $a$ is the scale factor.  The Hubble constant is $H=\dot{a}/a$, with $\dot{}\equiv d/dt$ where $t$ is the time. The evolution  equations are
\begin{eqnarray}
\label{evoleq}
\dot{\epsilon }_m &=&-3H\epsilon _m-\Gamma \epsilon _m, \nonumber \\
\dot{\epsilon }_r &=&-4H\epsilon _r+\Gamma \epsilon _m, \nonumber \\
H^2&=&{8\pi G \over 3} (\epsilon _r+\epsilon_m) 
\end{eqnarray}
After inflation  the fraction $x\equiv \epsilon _m/(\epsilon _m+\epsilon _r)=1$, and remains constant until it suddenly drops to zero when $H \sim \Gamma$. After that point $a\propto t^{1/2}$.  

\begin{figure}[tb] 
\centerline{\epsfxsize=9.0cm\epsffile{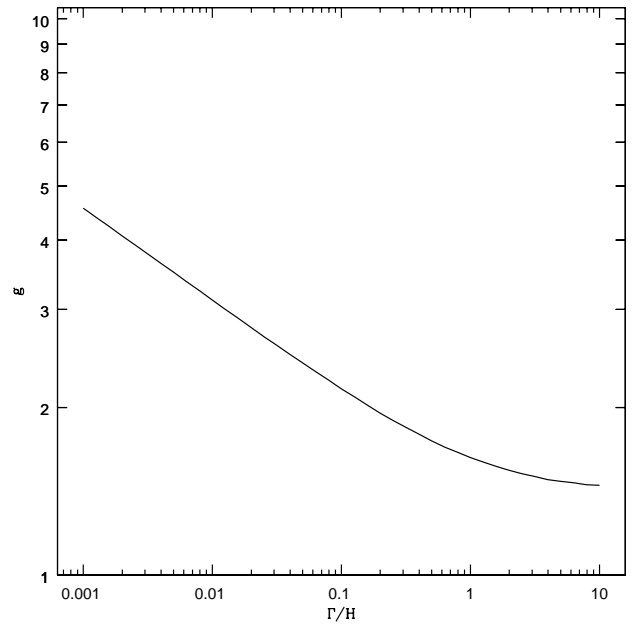}}
\smallskip
\caption{\label{fig1}\footnotesize%
Expansion factor normalization $g$ as a function of $\Gamma/H$ after reheating. The curve was obtained by solving equations (\ref{evoleq}) numerically starting with $\epsilon_r=0$ and Hubble expansion $H$ at the initial time.}
\end{figure}

The key point to understand is that universes with different values of $\Gamma$ will have expanded by different amounts by the time reheating in over. In figure \ref{fig1} we show the final scale factor as a function of $\Gamma$. We plot $\ln(g)=\ln(a t^{-1/2})$ vs. $\ln(\Gamma/H)$ where $H$ is the expansion rate at the end of inflation. The asymptote at $\Gamma/H \rightarrow 0$ corresponds to $g \propto \Gamma ^{-1/6}$,  a result that can be obtained analytically \cite{andrei}. 

After reheating is over the metric is $ds^2=-dt^2+g^2(\Gamma) t dx^2$ where $g^2(\Gamma)$ is the function plotted in figure \ref{fig1}. To obtain the density perturbations in our model we just need to promote $\Gamma$ to a function of position,
\beq{metric}
ds^2=-dt^2+g^2(\Gamma(x)) t dx^2. 
\eeq
This metric will be a solution of Einstein's equations if $\Gamma$ is sufficiently slowly varying, that is to say if $\Gamma$ changes on scales much larger than the horizon. Clearly this is a very good approximation as we are interested in changes of $\Gamma$ over scales that are of cosmological interest {\it today}. It is important to realize that we have not made an expansion in powers of $\delta \Gamma / \Gamma$, we have just assumed that the gradients are small. As a result we can use the above metric to calculate quantities to higher order in $\delta \Gamma /\Gamma$ as long as we do so when the variations of $\Gamma$ are on super-horizon scales. 

To make contact with observations we need to calculate the spatial curvature of surfaces of constant time. For the metric in equation (\ref{metric}) we obtain \cite{liddlelyth,mukh},
\begin{equation} \label{psi}
{\cal R} = {\delta g \over g} = - \alpha {\delta \Gamma \over \Gamma}.
\end{equation}
where $\alpha=-d\ln g /d\ln a$ is just minus the slope of the curve in figure $\ref{fig1}$ at the appropriate value of $\Gamma$.  This result was already obtained in \cite{dgz} by the standard perturbation theory methods.  In linear perturbation theory the variable $\cal R$ is related to the curvature of hypersurfaces of constant $t$ ($R^{(3)}$)  by (see for example \cite{liddlelyth})
\begin{equation}
R^{(3)} = - {4 \over a^2}  \nabla^2 \cal R.
\end{equation}
We will generalize this expression to include higher order terms later. 

A few things are worth noting. If after reheating the composition is the same everywhere, if we are dealing with adiabatic perturbations, the universe will expand by the same amount everywhere. As a result ratios of the scale factor between different places are  constant in time, and thus $\cal R$ is constant.   Certainly because the couplings and masses of particles are different in different places at reheating in this scenario there is the possibility of generating different ratios of abundances of particles as a function of location. Thus our scenario can create ``isocurvature" density perturbations as well. This is similar to what happens in the ``curvaton" scenario. Although this is a possibility, it is rather unlikely. Isocurvature perturbations could result if either there is varying decay into species out of equilibrium or if the decay of the  inflaton lead to  baryon asymmetry or lepton asymmetry fluctuations.

\section{Non-Gaussianities}\label{s2}


In this section we will discuss the level of non-Gaussianities that we can expect in this scenario. We will consider the curvature perturbations $\cal R$ and write,
\begin{eqnarray}
{\cal R} &=& {\cal R}_{g1} + f_{NL}^{{\cal R}_{g1}} ({\cal R}_{g1}^2 - \langle {\cal R}_{g1}^2 \rangle ) \nonumber \\ 
&+& f_{NL}^{{\cal R}_{g2}} ({\cal R}_{g2}^2 - \langle {\cal R}_{g2}^2 \rangle ) +  f_{NL}^{{\cal R}_{g1-g2}}{\cal R}_{g1}{\cal R}_{g2}+\cdots
\label{deffnls}
\end{eqnarray}
where ${\cal R}_{1g}$ and ${\cal R}_{2g}$ are Gaussian random fields and the $f_{NL}^{\cal R}$ control the level of non-Gaussianities. We have departed slightly from the notation of Komatsu \& Spergel \cite{komsper} by working with the curvature $\cal R$ rather than the gravitational potential because $\cal R$ is constant outside the horizon while the gravitational potential evolves whenever the effective equation of state of the universe changes. We have also allowed for the possibility that  more than one Gaussian field ${\cal R}_{1g}$ and ${\cal R}_{2g}$ enter into the quadratic corrections. As equation (\ref{gammaexp}) makes clear, this could happen if more than one field $\phi_i$ determined the decay rate of the inflaton. The first thing to point out is that the extra terms do not enter into the calculation of the three point function to lowest order in $f_{NL}$ so observationally they are not relevant. Higher order terms are suppressed by extra powers of ${\cal R}\sim 10^{-5}$.

There are several sources for the non-Gaussianities in ${\cal R}$.  The fluctuations $\delta \phi_i$ in equation (\ref{gammaexp}) could themselves be non-Gaussian. Even if those were Gaussian the relation between $\delta \phi_i$ and the decay rate $\Gamma$ in (\ref{gammaexp}) is non-linear. The fact that $\Gamma$ is by definition positive means that the relation {\it cannot} be purely linear, the size of those non-linearities however varies from model to model. Moreover as shown in  the previous section $g(\Gamma)$ is a non-linear function of $\Gamma$  which leads to additional non-Gaussianities. We will  consider the different cases in turn. 

\subsection{Gaussian $\delta \phi_i$ fluctuations}

We start by considering a simple example in which there is a field $\phi$ that sets the coupling of one of the channels through which the inflation can decay and that in addition there is an extra channel that does not fluctuate. This example was considered in \cite{dgz}. The decay rate can be written as,
\begin{eqnarray}
\Gamma&=&\Gamma_0 + \Gamma_1 ({ \phi \over \bar \phi})^2  \nonumber \\
&=&\Gamma_0 + \Gamma_1 ( 1 + {\delta \phi \over \bar \phi} )^2 , 
 \end{eqnarray}
so that 
\begin{eqnarray}
{\Gamma - \bar \Gamma \over \bar \Gamma}&=& {\Gamma_1 \over \bar \Gamma} ( 2 {\delta \phi \over \bar \phi} + {\delta \phi^2 - \langle \delta \phi^2 \rangle \over  \bar \phi^2}  )  \nonumber \\
\bar \Gamma  &=&\Gamma_0 + \Gamma_1 ( 1 + {\langle \delta \phi^2 \rangle \over \bar \phi^2}) . 
 \end{eqnarray}
The ratio $\Gamma_1 /  \bar \Gamma$ determines the fraction of the decay rate  that is controlled by $\phi$. 


We first consider the transformation between $\Gamma$ and $\cal R$ only to linear order,
\begin{equation}
{\cal R} = - \alpha {\Gamma_1 \over \bar \Gamma} ( 2 {\delta \phi \over \bar \phi} + {\delta \phi^2 - \langle \delta \phi^2 \rangle \over  \bar \phi^2}  ) 
\label{cal1}
\end{equation}
where $\alpha$ is less than $1/6$ (see figure \ref{fig1}).  We can rewrite equation (\ref{cal1}) as 
\begin{eqnarray}
{\cal R}&=& {\cal R}_g  + f_{NL}^{\cal R} ({\cal R}_g^2 - \langle {\cal R}_g^2 \rangle )  \nonumber \\
f_{NL}^{\cal R}  &=&  -{\bar \Gamma \over 4 \Gamma_1 \alpha } 
\label{cal2}
\end{eqnarray}

Equation (\ref{cal2}) shows that the smaller the efficiency of our mechanism of generating density perturbations, the larger the $f_{NL}^{\cal R}$. The inefficiency in the present example  comes from two sources, a small value of $\alpha$ which occurs when the decay rate of the inflaton in not much smaller than the Hubble constant at the end of inflation. The other source of inefficiency comes form the fact that the field only controls part of the channels through which the inflaton can decay. Again, the smaller the fraction of the decay rate controlled by $\phi$, the largest the non-Gaussianities. 

The origin of the relation between inefficiency and non-Gaussinanities is simple to understand.  If we keep the level of fluctuations in $\cal R$ fixed, which me must to explain the observed level of density perturbations, as we decrease the efficiency we must increase the level of $\delta \phi /\bar \phi$. As a result quadratic  terms in  $\delta \phi / \bar \phi$ become more important relative to the linear term and thus  the non-Gaussianity increases. As this arguments makes clear, the relation between inefficiency and non-Gaussianity  is more general than the particular example we have chosen for the form of the decay rate. 

We now need to consider the fact that the relation between the decay rate and the final curvature is non-linear, that is the final value of the scale factor is a non-linear function of the $\Gamma$ as figure \ref{fig1} shows. To be more specific we will calculate the curvature of the hypersurfaces of constant time in the metric of equation (\ref{metric}). We can do so without doing a linear approximation in $\delta \Gamma / \Gamma$ although we are calculating only the leading contribution in an expansion in terms of the horizon scale over the scale of variation of $\Gamma$. We get,
\beq{r3}
R^{(3)}=-{4 \over a^2} [\nabla^2 \ln g - ||\nabla \ln g||^2],
\eeq
where $a=g(x) t^{1/2}$. The first term is the direct generalization of the linear perturbation formula, with ${\cal R} = \ln (g /\bar g)$. As it should be only spatial changes the logarithm of $g$ signal curvature perturbation. 

We are interested in calculating the curvature to second order in $\delta \Gamma/\Gamma$.  There will be two contributions, one from expanding the $\ln g$ to second order and the other from the second term in equation (\ref{r3}).  These second term arises because General Relativity itself is a non-linear theory. We will comment on these shortly. For the moment we focus on the first term. With the definition ${\cal R} = \ln(g /\bar g)$ we have,
\beq{nonlR}
{\cal R}= -{\alpha} {\delta \Gamma \over \bar \Gamma} + {\alpha+\beta \over 2}   {\delta \Gamma ^2 - \langle  \delta \Gamma ^2 \rangle  \over \bar \Gamma}  +\cdots
\eeq
where $\alpha=-d\ln a / d\ln \Gamma$, the slope of the curve in figure \ref{fig1} and $\beta$ is the curvature $\beta=d^2\ln a / d\ln \Gamma^2$.  Thus the fact that the relation between $\Gamma$ and total expansion is non-linear adds an additional contribution to $f_{NL}^{\cal R}$,
\beq{nonlr}
f_{NL}^{\cal R} = {\alpha+\beta \over 2 \alpha^2} 
\eeq
This contribution is at least equal to $f_{NL}^{\cal R}=3$ and increases as the process becomes more inefficient $(\alpha < 1/6)$  because $\delta \Gamma /\bar \Gamma$ increases to produce the same level of curvature fluctuations. Note however  that these non-Gaussianities do not include the extra enhancement factor $\bar \Gamma/\Gamma_1$. The point is that the inefficiency in the translation between $\delta\Gamma /\bar \Gamma$ and $\delta \phi/\bar\phi$ makes $\delta \phi / \bar \phi$ larger and thus terms quadratic in $\delta \phi / \bar \phi$ lead to larger non-Gaussianities than terms arising from $\delta \Gamma / \bar \Gamma$ corrections. 

We now comment on the second term in equation (\ref{r3}). It contributes an additional effect that leads to an $f_{NL}^{\cal R} \sim -1$, thus it is subdominant to what comes from the first term and is independent of $\alpha$. Moreover as the modes enter the horizon additional non-Gaussianities from non-linearities caused by term we have neglected in  Einstein's equations will develop. Perhaps more importantly one should solve for the observables (such as the CMB anisotropies) to second order in $\cal R$ to catch all the terms that contribute at the level of $f_{NL}^{\cal R} \sim 1$.  $R^{(3)}$ may not be the quantity directly relevant to observations. It is beyond the scope of this paper to characterize all these terms or to extend the usual linear treatment of CMB anisotropies  to include them. {\it What we have calculated should be interpreted as corrections in addition to what GR already gives}. 

Finally we want to comment on the possibility that the quadratic corrections to the Gaussian curvature are not proportional to the linear contribution squared. In our scenario this happens trivially when more than one light field determines $\Gamma$. For example if
\begin{eqnarray}
\Gamma&=&\Gamma_0 + \Gamma_1 ({ \phi_1 \over \bar \phi_1})^2  + \Gamma_2 ({ \phi_2 \over \bar \phi_2})^2
 \end{eqnarray}
then there are two different Gaussian fields that enter in the quadratic corrections,
\beqa{rg12}
{\cal R}_{g1}&=& {2 \Gamma_1 \over \bar \Gamma} {\delta\phi_1 \over \bar \phi_1} +
{2 \Gamma_2 \over \bar \Gamma} {\delta\phi_2 \over \bar \phi_2}  \nonumber \\
&& \nonumber \\
{\cal R}_{g2}&=& {2 \Gamma_2 \over \bar \Gamma} {\delta\phi_1 \over \bar \phi_1} -
{2 \Gamma_1 \over \bar \Gamma} {\delta\phi_2 \over \bar \phi_2}.
\eeqa
The amplitudes of the non-linear terms defined in equation (\ref{deffnls}) are,
\beqa{fnls}
f_{NL}^{{\cal R}_{g1}} &=& {[(\Gamma_1/\bar \Gamma)^2+(\Gamma_2/\bar \Gamma)^2]^{-1/2}\over 4} (\sin^3\theta+\cos^3 \theta) \\
f_{NL}^{{\cal R}_{g2}} &=&  {[(\Gamma_1/\bar \Gamma)^2+(\Gamma_2/\bar \Gamma)^2]^{-1/2}\over 4} \sin\theta \cos\theta (\sin\theta+\cos \theta) \nonumber \\
f_{NL}^{{\cal R}_{g1-g2}} &=&  {[(\Gamma_1/\bar \Gamma)^2+(\Gamma_2/\bar \Gamma)^2]^{-1/2}\over 2}  \sin\theta \cos\theta (\sin\theta-\cos \theta),\nonumber 
\eeqa
with $\cos\theta = \Gamma_1/[\Gamma_1^2+\Gamma_1^2]^{1/2}$ and $\sin\theta = \Gamma_2/[\Gamma_1^2+\Gamma_1^2]^{1/2}$. Other than the slight redefinition of $f_{NL}^{{\cal R}_{g1}} $ these additional terms are unimportant if we are only interested in calculating effects to first order in  the $f_{NL}$s. The other terms will only enter at higher orders and will be thus suppressed by extra powers of ${\cal R}\sim 10^{-5}$. Generically then, modifications of the standard $f_{NL}$ form, ${\cal R} ={\cal R}_{g} + f_{NL}^{{\cal R}} ({\cal R}_{g}^2 - \langle {\cal R}_{g}^2 \rangle )$, are what is expected in our scenario. It turns out this remains true even when self couplings of the field are considered.  

\subsection{non-Gaussian $\delta \phi_i$ fluctuations}

In this section we calculate the non-Gaussianities in the $\delta \phi$ fluctuations if the potential for $\phi$ is not quadratic. We will follow \cite{maldacena}, although our situation is much simpler because the field makes a negligible contribution to the energy density so we do not need to take into account perturbations in the metric .
The action for $\delta \phi$ is, 
\beqa{action}
S &=& \int a^4 \ d^3 x d \eta \ [ {(\delta \phi^{\prime})^2 - (\nabla \delta \phi)^2  \over 2 a^2 } \nonumber \\ 
& &  - {V^{(2)} \over 2 !}  \delta \phi^2  - {V^{(3)} \over 3 !}  \delta \phi^3  - {V^{(4)} \over 4 !}  \delta \phi^4 - \cdots ].
\eeqa
We denote $V^{(k)}$ the $k$th derivative of the potential. 

We quantize the field in the usual way to obtain,
\beq{quant}
\hat {\delta \phi} = \int {d^3 \k \over (2 \pi)^3} [ f(k,\eta) e^{- i \k \cdot \x} a_{\k} + f^*(k,\eta) e^{i \k \cdot \x} a^\dagger_{\k} ]
\eeq
where $[a_{\k_1}, a^\dagger_{\k_2}] = (2\pi)^3 \delta^D(\k_1-\k_2)$ and $f(k,\eta) $  are the mode functions. For the massless field in de-Sitter space ($V^{(2)}=0$ and $H$ constant),
\beq{f}
f(k,\eta)= {H \over \sqrt{2 k^3} } (1-i k \eta ) e^{i k \eta}
\eeq

We want to calculate the N-point function $g_N(1,2, \cdots N)$ for $N$ points at the time of rehating,
\beqa{npoint1}  
g_N(1,2, \cdots N) &=&\langle 0 | U^{-1} \hat \delta \phi(1)\hat \delta \phi(2) \cdots \hat \delta \phi(N) U | 0 \rangle \nonumber \\
U&=&\exp{[-i \int_{t_0}^{t_{reh}} \hat H_{int} \ dt] }
\eeqa
where $t_{reh}$ is the time of rehating, $t_0$ is some early initial time and $\hat H_{int}= {V^{(3)} \over 3 !}\delta \phi^3  + {V^{(4)} \over 4 !}  \delta \phi^4 + \cdots$ is the Hamiltonian in the interaction picture. To lowest order in $\hat H_{int}$ this gives,
\beqa{npoint2}  
g_N(1,\cdots N) &=&-i  \int dt \langle 0 | [ \hat \delta \phi(1)\hat \delta \phi(2) \cdots \hat \delta \phi(N),  \hat H_{int} ] | 0\rangle \nonumber
\eeqa

The N-point function is most easily expressed in Fourier space. For a massless field in de-Sitter we obtain,
\beqa{npointF}
\tilde g (\k_1,\cdots \k_N)&&= (2\pi)^3  \delta^D(\k_1+\cdots +\k_N) \times \nonumber \\
&&  V^{(N)} { H^{2N} k_t^3 \over \prod_i 2 k_i^3} I_N{(k_1,k_2,\cdots k_N})  \\
I_N= \int_{-\infty}^{\eta_{reh}} {d \eta \over k_t^3 \eta^4} && \ 2 {\rm Re}[ - i (1-i k_1 \eta)  \cdots (1-i k_N \eta) e^{i k_t \eta}]   \nonumber   
\eeqa
with $k_t = k_1 + k_2 + \cdots +k_N$ and $\eta_{reh}$ is the conformal time of rehating. Following \cite{maldacena} to project onto the vacuum of the interacting theory  we deform the time integration so it has some evolution in euclidean time. We explicitly give the results for the 3- and 4-point functions,
\beqa{34point}
I_{3}&=& {8 \over 9} - {\sum_{i<j} 2 k_i k_j \over k_t^2 } - {1 \over 3} (\gamma+N_{k_t})   
{\sum_{i}  2 k_i^3 \over k_t^3 } \\
I_{4}&=& {8 \over 9} - {\sum_{i<j} 2 k_i k_j \over k_t^2 } + 2 {\prod_i k_i \over k_t^4} - {1 \over 3} (\gamma+N_{k_t})   
{\sum_{i}  2 k_i^3 \over k_t^3 } \ \nonumber 
\eeqa
where $\gamma= 0.577216 $ is the Euler's constant and $N_{k_t}=\ln(|k_t \eta_{reh}|)$ is the number of e-foldings from the time  $k_t$ crossed the horizon ($k_t \eta = -1$) to the time of reheating. 

The value of $I_N$ is dominated by the term proportional to $N_{k_t}$ because the modes that we observe left the horizon roughly 60 e-foldings before the end of inflation.
This has a very simple interpretation. All terms in  equation (\ref{34point}) are of order one except for that proportional from  $N_{k_t}$. They represent the effect of the self-coupling on the modes as they are being generated, at horizon crossing during inflation. The terms proportional to  $N_{k_t}$ however originates from  the classical evolution of the field outside the horizon. To see this in a concrete example lets consider the corrections induced by classical outside the horizon evolution of a field with a small third and fourth order coupling. We will write $\delta\phi(\x)=\delta\phi_{(0)}(\x)+\delta\phi_{(1)}(\x) + \cdots $ where $\delta\phi_{(0)}$(\x)  denotes the Gaussian fluctuations that are produced in the absence of these couplings,  $\delta\phi_{(1)}(\x)$ the first order corrections, etc. We can neglect spatial derivatives and assume the field is slowly rolling so that, 
\beq{calss1}
3 H \dot{ \delta\phi}_{(1)} = - {V^{(3)} \over 2 !}  \delta\phi_{(0)}^2 - {V^{(4)} \over 3 !}  \delta\phi_{(0)}^3. 
\eeq
Thus in de-Sitter space ($H$ constant)
\beq{class2}
\delta\phi_{(1)} =- {N \over 3 H^2 } [ {V^{(3)} \over 2 !}  \delta\phi_{(0)}^2 + {V^{(4)} \over 3 !}  \delta\phi_{(0)}^3]. 
\eeq
where we have introduced the number of e-foldings $N=H \Delta t$. The three and four point functions implied by these expressions are exactly the same as those corresponding to the terms proportional to $N_{k_t}$ in equation (\ref{34point}). A discussion of the non-Gaussianites generated by outside the horizon classical evolution of a scalar field, in the context of a multicomponent inflationary model can be found in \cite{bernardeau2}.

The physics  is very simple, the self-couplings of the field creates higher order correlations in two different ways. It modifies the the statistics of the fluctuations themselves but it also changes the Gaussian part of the fluctuations as the field evolves outside the horizon. Because we are considering the evolution outside the horizon it is local in real space, meaning that it  leads to  non-Gaussianities of the form,
\beq{calss3}
\delta\phi =\delta\phi_{(0)}+ a_2 \delta\phi_{(0)}^2+ a_3 \delta\phi_{(0)}^3 + \cdots
\eeq
There are deviations from this ``local" approximation coming from the fact that $N_{k_t}$ varies across the scales we observed, it  changes by roughly 10 \% between the CMB scales and smaller scales that one could access by studying large scale structure. There are also corrections of order $1/ N_{k_t}$ due to the non-Gaussianities produced at the time the fluctuations crossed the horizon during inflation. Those have different configurational dependance. 

We should also point out that the non-linearities keep acting all the way until reheating, even after the universe stopped inflating. During the period of matter domination, what determines the growth is,
\beq{nreh}
\int {dt \over H} \approx {3 \over 2 H_{reh}^2} =  {3 \over 2 \Gamma^2},
\eeq
 where we have just taken $H\sim 2/3 t$ during the period between the end of inflation and reheating and used $H_{reh}=\Gamma$. Thus we should replace the number of e-foldings in equation (\ref{class2}) by an effective number $\tilde N_{k_t} = N_{k_t} + 3/2 (H/\Gamma)^2$ with $H$ the Hubble constant during inflation.

We will concentrate on the three point function because unless $\phi$ is fluctuating very close to the origin, so that $\delta \phi /\bar \phi \sim 1$ the four point function will be smaller. In fact one naturally expects $\langle {\cal R}^4 \rangle /\langle {\cal R}^3 \rangle \langle {\cal R}^2 \rangle^{1/2} \sim \delta \phi /\bar \phi$. For our simple one field example the contribution to $f_{NL}^{\cal R}$ 
due to non-linearities in  the potential is, 
\beq{fnl1}
f_{NL}^{\cal R}  = -  {\bar \Gamma \over \Gamma_1 \alpha } { V^{(3)} \bar \phi \over 3 H^2}\tilde  N_{k_t}
\eeq

To make an estimate of the size of these non-Gaussian contribution we will relate $V^{(3)}$ to 
$V^{(2)}=m^2$ and rewrite, 
\beq{fnl2}
f_{NL}^{\cal R}  = -  {\bar \Gamma \over \Gamma_1 \alpha } \ ({ V^{(3)} \bar \phi \over m^2 }) \ ({ 2 m^2 \over 3 H^2}) \  {\tilde N_{k_t} \over 2}.
\eeq
We now notice the in the presence of a mass term the evolution of the modes outside the horizon  ($3 H \dot{ \delta\phi}_{(1)} = - m^2  \delta\phi_{(0)}^2$) creates as scale dependance of the power spectrum because larger wavelengths that came out the horizon earlier had a longer time to evolve and thus have a smaller amplitude at reheating. This is exactly the same as what happens in the curvaton scenario \cite{lyth}.  
The power spectrum spectrum gets a blue tilt from this effect,
\beq{tilt}
(n-1)^{\rm mass} = {2\over 3} \ {m^2 \over  H^2} 
\eeq
so that,
\beq{fnl3}
f_{NL}^{\cal R}  = -  {\bar \Gamma \over \Gamma_1 \alpha } \ ({ V^{(3)} \bar \phi \over m^2 }) \ 
(n-1)^{\rm mass}  \  {\tilde N_{k_t} \over 2}.
\eeq
Departures from scale invariance are observed to be small, that is $n-1$ should be smaller than few $\times \ 10^{-2}$ while $\tilde N_{k_t}$ is of order 60-100.   If the factor $m^2/ 3 H^2 \times \tilde N_{k_t} \sim 1$ and $V^{(3)} \bar \phi / m^2 \sim 1$ then the non-Gaussianities produced by the self couplings are of the same order than those produced by the non-linear relation between the fluctuating field and the decay rate. To put it another way, even though $V(\phi)$ does not have to satisfy slow-roll conditions, if the spectrum of perturbations in nearly scale invariant then the non-Gaussainities are not expected to be larger than current observational constraints. 

We can consider a simple example $V(\phi)= \lambda \phi^4 /4!$ and assume that inflation has lasted long enough before the scale of our present horizon crossed the horizon during inflation, that the field $\phi$ has rolled to the point in the potential where the change during one e-folding in the classical trajectory equals the quantum fluctuations $H/2\pi$. This point corresponds to $\bar \phi \sim  \lambda^{-1/3} H$. In that case $(n-1)^{\rm mass} \sim \lambda^{1/3}$. Thus constraints on the spectral index already put $\lambda <  10^{-6}$. Moreover $\delta \phi/\bar \phi \sim \lambda^{1/3} /2\pi$ so to match the observed ${\cal R} \sim 4  \ \times \ 10^{-5}$ we need $ {\Gamma_1 \alpha/\bar \Gamma } \sim 10^{-4}/\lambda^{1/3}$.  
The non-Gaussianity parameter is then, 
\beq{fnl4}
f_{NL}^{\cal R}  \sim -   3 \ 10^{3}  \lambda^{2/3}  {\tilde N_{k_t} }.
\eeq
So for example if $\lambda^{1/3} \sim 10^{-2}$ and $\tilde N_{k_t} \sim 100$ we get $f_{NL}^{\cal R}  \sim -30$ very close to the limit set by WMAP. 

\section{Summary}\label{s3}


We have discussed the different sources of non-Gaussianities in the scenario introduced in \cite{dgz}.  In this section we summarize the results and make contact with observations. In most of the observational studies so far, in particular in the latest constraints form WMAP \cite{komatsu} the analysis was done in terms of the non-Gaussianities in the gravitational potential during the matter dominated era, $\Phi = - 3/5 {\cal R}$. Constraints were set on  $f_{NL}^{\Phi}$ defined by,
\begin{equation}
{\Phi} = {\Phi}_{g}  + f_{NL}^{\Phi} ({\Phi}_{g}^2 - \langle {\Phi}_{g}^2 \rangle ) + \cdots.
\end{equation}
Thus $f_{NL}^{\Phi}=-5/3 f_{NL}^{\cal R}$. WMAP constrained 
$-58 < f_{NL}^{\Phi}< 134$ at 95 \% confidence \cite{komatsu}. 

Non-Gaussainities in this scenario can have several sources. There are contributions that are proportional to $\bar \Gamma /\Gamma_1 \alpha$. These come from corrections that are directly proportional to $\delta \phi/\bar \phi$, either from non-linearities in the relation between $\delta \Gamma /\bar \Gamma$ and $\delta \phi / \bar \phi$ or intrinsic non-Gaussianities due to a cubic self interaction of the field. In the simple examples we have presented, the first contribution is 
$ f_{NL}^{\Phi} =5/12\ (\bar \Gamma /\Gamma_1 \alpha)$. The WMAP constraints already imply that $\Gamma_1 /  \bar \Gamma > 0.1$.   The contribution from the self couplings, equation (\ref{fnl1}), combined with the WMAP constraints imply that  ${ \tilde N_{k_t}  V^{(3)} / H } < 0.1$. 

There are non-Gaussianities coming from the non-linear relation between $\cal R$ and $\delta \Gamma /\bar \Gamma$. These grow as the generation of density perturbations becomes more inefficient, that is as $\bar \Gamma / H$ grows but they are not proportional to $\bar \Gamma / \Gamma_1$. In the limit $\bar \Gamma << H$  this contribution is $ f_{NL}^{\Phi} = - 5$ and is larger if $\alpha < 1/6$. 

Finally,  all these contributions should be interpreted as additions to the $ f_{NL}^{\Phi} \sim 1$ coming from second order GR effects which we have not considered here.  For  effects produced when modes are crossing the horizon (either on the way out or in)  spatial gradients will not be negligible so the configuration dependance of the three point function  will be different than in  the simple $f_{NL}$ model. These corrections are independent of the efficiency of our  the mechanism to translate $\delta \phi /\bar \phi $ into $\cal R$ and thus are expected to be the smallest. 


A few general conclusions of our study are worth emphasizing. First as noted above, inefficiencies in the translation between field fluctuations and curvature perturbations increase the expected non-Gaussianities.  Second, the non-Gaussianities in our model are mainly generated when the modes of interest are outside the horizon. As a result it is natural to look for departures from Gaussianity that are ``local" in real space, meaning of the form ${\cal R} = {\cal R}_{g} + f_{NL}^{\cal R} ({\cal R}_{g}^2 - \langle {\cal R}_{g}^2 \rangle ) + \cdots$. If the decay rate is determined by more that one fluctuating field there could be additional corrections at the second order in $\cal R$, but these are irrelevant if one is interested only in N-point functions to leading order  in $f_{NL}$ (the actual expansion parameter is $f_{NL}^{\cal R} {\cal R}$ so higher order terms are very small).   The ``local " approximation should be excellent in these models. Third because our mechanism is in some sense somewhat inefficient ($\alpha < 1/6$, $\Gamma_1/\bar \Gamma < 1$), $f_{NL}\sim$ few is guaranteed in our scenario even if the fluctuations in the fields are intrinsically Gaussian. This last point is of particular interest because $f_{NL}\sim$ few corresponds to the expected sensitivity of future CMB experimets \cite{komsper} and the standard inflationary scenario predicts corrections much smaller than these \cite{maldacena}. 
If the efficiency of the mechanism is decreased or if self interactions are important, the level of non-Gaussianities can be increased to saturate the current upper limits. However the fact that the spectrum of perturbations is close to scale invariant, implies that even in the presence of self-couplings the natural  level of non-Gaussianities is somewhat below the current limit although not far from it. 

\bigskip
{\bf Acknowledgments}
We thank Nima Arkani-Hammed, Paolo Creminelli Gia Dvali, Andrei Gruzinov and Juan Martin Maldacena for useful discussions. M.Z. is supported by the David and Lucile  Packard Foundation and NSF.

\ed

\bibitem{reheating} L. Kofman, A. Linde, A. A. Starobinsky, PRD, 56, 3258 (1997)

To obtain an analytic solution it convenient to describe reheating by one third-order equation
\begin{equation} \label{a3p}
a'''=-\Gamma aa'',
\end{equation}
where $'=ad/dt\equiv d/d\eta $ is the conformal time derivative. This is a non-linear equation for $a$ so the value of $a$ after reheating will be a non-linear function of $\Gamma$.

Before reheating, $a=C_mt^{2/3}$. Given the constant $C_m$ we can write the solution of eq. (\ref{a3p}) in the following form 
\begin{equation}
a=AF(A\Gamma \eta).
\end{equation}
Here $F(\eta)$ denotes the solution of
\begin{equation}
F'''=-FF''.
\end{equation}

for simplicity we now consider the limit of $\Gamma <<  H$ initially so that we have,
with  $F\rightarrow \eta ^2$ for $\eta\rightarrow 0$, and $A= (3\Gamma )^{-2/3}C_m$.
For large $\eta$, $F(\eta ) \propto \eta $, giving the state of the universe after reheating ends. For large $t$, we get $a=C_rt^{1/2}$, with
\begin{equation} \label{result}
C_r\propto C_m\Gamma ^{-1/6}
\end{equation}
Now assume that $\Gamma (x)$ is actually a slowly (super-horizon) varying function of position. Neglecting metric perturbations from inflation means taking $C_m$ that does not depend on $x$. Eq.(\ref{result}) remains valid to zeroth order in spatial gradients, giving the post-reheating metric (also to zeroth order in spatial gradients) of the form
\begin{equation} \label{metric}
ds^2=dt^2-C_r^2(x)tdx^2,
\end{equation}
where
\begin{equation} \label{gamma}
C_r(x)\propto \Gamma ^{-1/6}(x).
\end{equation}

\section{Examples}

Should we give a few examples here?
Possible nice things to show are examples different from $\Gamma \propto \phi^2$ 
examples where $\phi$ has a potential different from quadratic, examples of cases when more than one field are relavant. 

For simplicity we will give results for the case in which $\bar \Gamma << H$ at the end of inflation as this limit can be solved analytically. The curvature fluctuations in these case are given by, 
\begin{eqnarray}
{\cal R} &=& {\Gamma(x)^{-1/6} - \bar \Gamma^{-1/6} \over \bar \Gamma^{-1/6} } \nonumber \\
&=& -{1 \over 6} {\delta \Gamma \over \bar \Gamma} + {7 \over 72}   {\delta \Gamma ^2 - \langle  \delta \Gamma ^2 \rangle  \over \bar \Gamma} 
\end{eqnarray}
{\bf define curvature !!!!!}